\documentclass[a4paper,fleqn,usenatbib]{mnras}
\usepackage{txfonts}

\usepackage[T1]{fontenc}
\usepackage{ae,aecompl}

\usepackage{graphicx}	
\usepackage{amssymb}	

\title[The importance of background decontamination]{No evidence for younger stellar generations within the intermediate age massive clusters NGC 1783, NGC 1696 and NGC 411}
\author[I. Cabrera-Ziri et al.]{I. Cabrera-Ziri$^{1,2}$\thanks{ICZ: icabrera@eso.org}, F. Niederhofer$^{3,4,5}$, N. Bastian$^{2}$, M. Rejkuba$^{1,3}$, E. Balbinot$^6$,\linebreak \newauthor W. E. Kerzendorf$^{1}$, S. S. Larsen$^{7}$, A. D. Mackey$^{8}$, E. Dalessandro$^{9}$, A. Mucciarelli$^{9}$,\linebreak \newauthor C. Charbonnel$^{10,11}$, M. Hilker$^{1}$, M. Gieles$^6$, V. H\'enault-Brunet$^6$
\\
$^{1}$ European Southern Observatory, Karl-Schwarzschild-Stra{\ss}e 2, D-85748 Garching bei M{\"u}nchen, Germany\\
$^{2}$ Astrophysics Research Institute, Liverpool John Moores University, 146 Brownlow Hill, Liverpool L3 5RF, UK\\
$^{3}$ Excellence Cluster Origin and Structure of the Universe, Boltzmannstr. 2, D-85748 Garching bei M{\"u}nchen, Germany\\
$^{4}$ Universit{\"a}ts-Sternwarte M{\"u}nchen, Scheinerstra{\ss}e 1, D-81679 M{\"u}nchen, Germany\\
$^{5}$ Space Telescope Science Institute, 3700 San Martin Drive, Baltimore, MD 21218, USA\\
$^{6}$ Department of Physics, University of Surrey, Guilford GU2 7XH, UK\\
$^{7}$ Department of Astrophysics / IMAPP, Radboud University, PO Box 9010, NL-6500 GL Nijmegen, the Netherlands\\
$^{8}$ Research School of Astronomy and Astrophysics, Australian National University, Canberra, ACT 2611, Australia\\
$^{9}$ Dipartimento di Fisica \& Astronomia, Universit\`a degli Studi di Bologna, Viale Berti Pichat, 6/2 -40127, Bologna, Italy\\
$^{10}$ Department of Astronomy, University of Geneva, Chemin des Maillettes 51, 1290, Versoix, Switzerland\\
$^{11}$ IRAP, UMR 5277 CNRS and Universit\'e de Toulouse, 14 Av. E. Belin, 31400, Toulouse, France
}

\date{Accepted XXX. Received YYY; in original form ZZZ}

\pubyear{2016}

\begin{document}
\label{firstpage}
\pagerange{\pageref{firstpage}--\pageref{lastpage}}
\maketitle

\begin{abstract}
Recently, \cite{Li16} claimed to have found evidence for multiple generations of stars in the intermediate age clusters NGC 1783, NGC 1696 and NGC 411 in the Large and Small Magellanic Clouds (LMC/SMC). Here we show that these young stellar populations are present in the field regions around these clusters and are not likely associated with the clusters themselves.  Using the same datasets, we find that the background subtraction method adopted by the authors does not adequately remove contaminating stars in the small number 
 Poisson limit. Hence, we conclude that their results do not provide evidence of young generations of stars within these clusters.

\end{abstract}


\section{Introduction}
\label{sec:intro}

The peculiar abundance patterns found in globular clusters, and the complex colour-magnitude diagrams (CMDs) of young and intermediate age clusters in the Magellanic Clouds have turned the study of multiple stellar populations (MPs) in clusters into a very active research field. Often, these MPs have been hypothesised to be associated with different stellar generations -- i.e., distinct epochs of star formation (e.g. \citealt{Mackey08} in the context of intermediate age clusters in the Magellanic Clouds). However, definitive evidence for multiple stellar generations within stellar cluster remains elusive to date\footnote{With the notable exception of nuclear stars clusters e.g. \cite{Walcher06}.} (cf. \citealt{Li14}; \citealt{cz14,cz16a,Niederhofer15}).

Recently, Li et al.~(2016; hereafter L16), studied the CMDs of three intermediate age ($\sim 1.5$ Gyr) clusters in the LMC/SMC (NGC 1783, NGC 1696 and NGC 411), and claimed to have found young (few hundred Myr) populations of stars in each cluster. The authors interpreted these results as a ``smoking gun'' of a recent star formation burst within these clusters.

In this paper, we use L16 photometric catalogs, and find that the young populations on the CMDs of these three clusters are also present in the CMDs of field regions around the clusters, challenging the associations of the young populations with these clusters.

\section{Densities and luminosity functions of the young stars}
\label{sec:dens}

\begin{table*}
\caption{Number ($N$ and $N_F$) and surface density ($\Sigma_N$ and $\Sigma_{N_F}$) of young stars in the cluster region (before decontamination) and the reference field.}
\begin{center}
\begin{tabular}{cccccc}
\hline
NGC & $N$ & $N_F$ & $\Sigma_N$ & $\Sigma_{N_F}$ & $\Sigma_{N_F} - \Sigma_N$\\
& (stars) & (stars) & ($\times10^{-3}$ stars/arcsec$^2$) & ($\times10^{-3}$ stars/arcsec$^2$) &($\times10^{-3}$ stars/arcsec$^2$)\\
\hline
1783 & 167 & 311 & $5.32(\pm0.41)$ & $7.77(\pm0.44)$ & $-2.49 (\pm0.60)$\\
1696 & 148 & 55 & $13.09(\pm1.08)$ & $9.17(\pm1.24)$& $3.92 (\pm1.64)$\\
411 & 86 & 34 & $10.95(\pm 1.18)$ & $7.59(\pm1.30)$& $3.36 (\pm1.76)$\\
\hline
\end{tabular}
\end{center}
\label{tab1}
\end{table*}%

\begin{figure}
\centering
\includegraphics[width= 72mm]{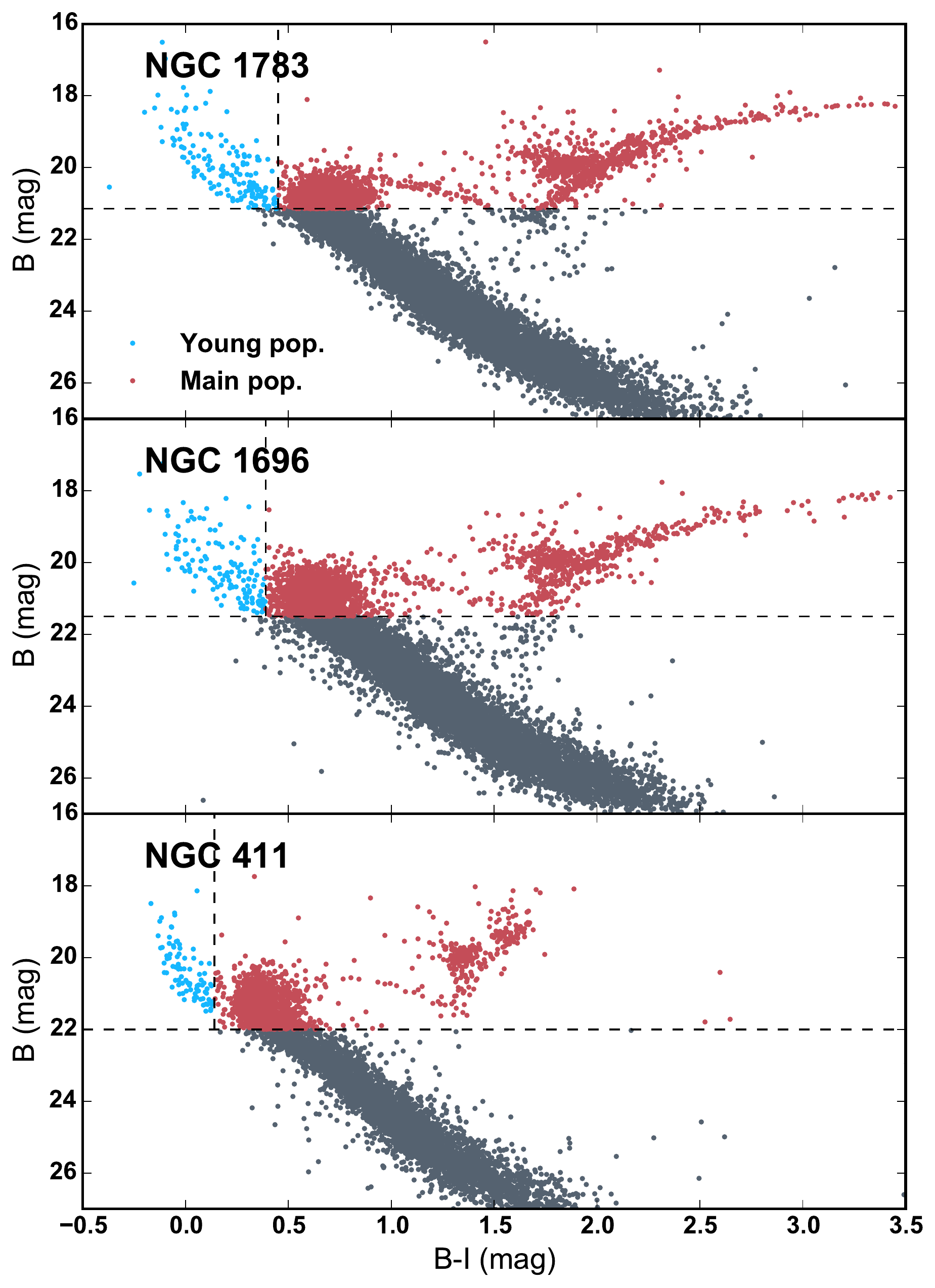}
\caption{Selection of young (blue dots) and main ($\sim1.5$ Gyr old -- red dots) populations from the CMDs (before decontamination) for the luminosity functions (cf. Fig. \ref{LFs}) and radial profiles (cf. Fig. \ref{RPs}).}
\label{young_old}
\end{figure}

We compared the density of young stars in the cluster (i.e. inner two core radii, as defined by L16) and in the reference field region, which was chosen to be the same as in L16. This comparison was done before applying any decontamination and the definition of the young sequences in the CMD was selected to be
$B\mbox{(mag)}<\{21.25,21.50,22.00\}$ and $B-I\mbox{(mag)}<\{0.45,0.39,0.14\}$ for clusters NGC 1783, NGC 1696 and NGC 411 respectively (cf. Fig. \ref{young_old}). In Table \ref{tab1} we report the number and (average) surface density of the young stars in the cluster and reference field region for each case. In brackets we show the standard deviation of the densities calculated $\sqrt{N}/A_C$ and $\sqrt{N_F}/A_{F}$ where $A_C$ and $A_F$ are the solid angle (area) of the cluster and reference field region.

The last column of the table shows the difference between surface densities. For NGC 1783, we find that the reference field contains significantly ($\sim4\sigma$) more of these young stars per unit area than the cluster. While for NGC 1696 and NGC 411 both the densities of young stars in the cluster is 0 within $\sim2.3\sigma$ and $\sim1.9\sigma$ respectively. From this, we find no significant overdensities of young stars in the cluster regions with respect to the reference field.

\begin{figure}
\centering
\includegraphics[width= 72mm]{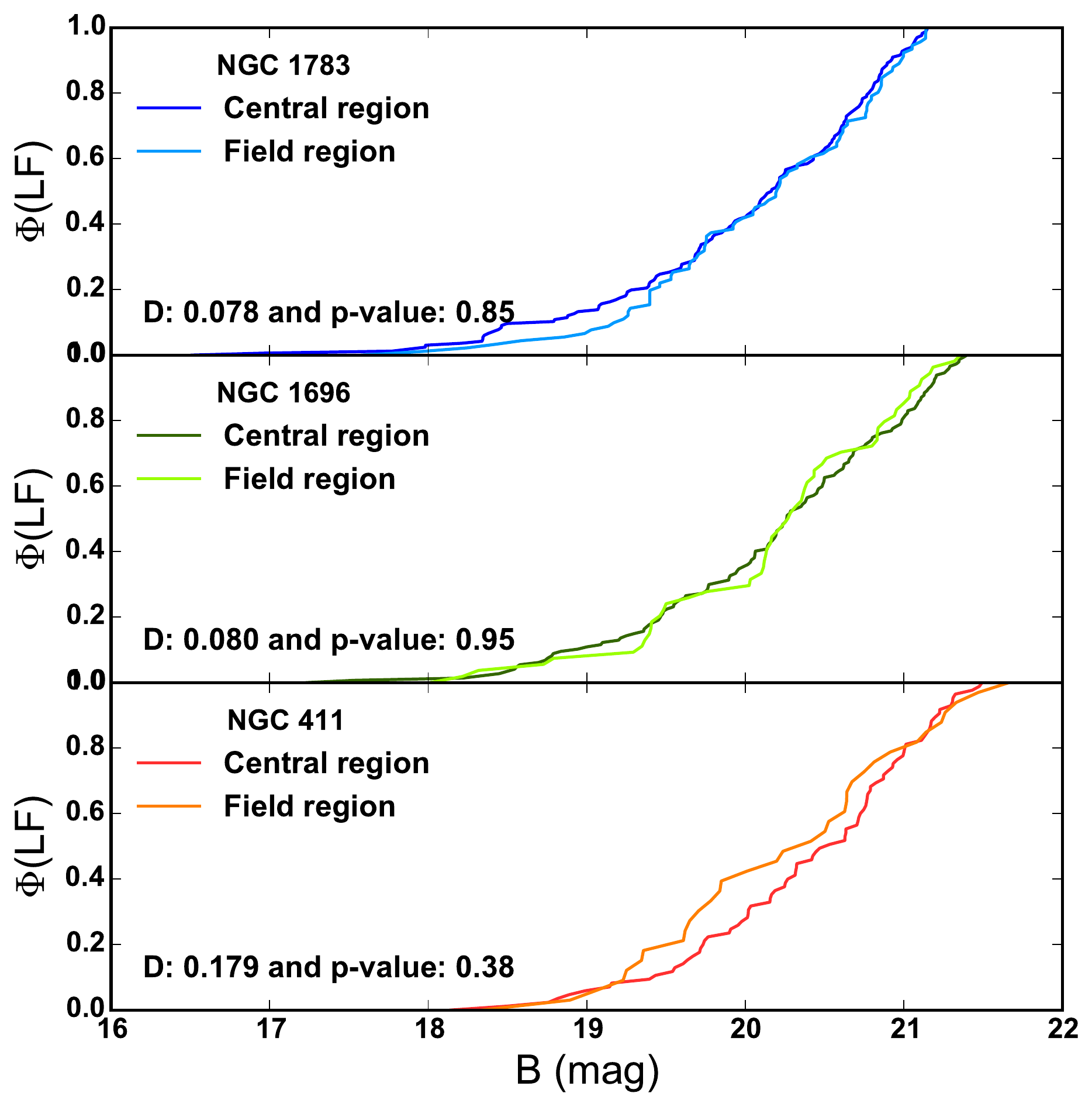}
\caption{Cumulative LFs of the young populations in the clusters and field regions. In each panel we report the KS statistic, D, and p-value from our analysis of the LFs of the young populations from the field and cluster regions. We find no significant differences between them.}
\label{LFs}
\end{figure}

Additionally, we calculated the luminosity functions (LFs) of these young populations in the field region and compared them with the LFs of the young populations in the cluster region (also before applying any decontamination). The cumulative LFs of the young populations from clusters and field regions are shown in Fig. 2. They are very similar in all cases.
For every cluster we applied a KS test to compare the LFs of both regions, the results are shown in Fig. \ref{LFs} as well. From the KS test we can say that the LFs of the young populations in the field regions around the clusters, and the LFs of the stars within the clusters, do not show any significant difference.

This casts doubts on the association of these young populations with the clusters themselves as 1) there are no obvious overdensities of young stars in the cluster with respect to the reference fields, and 2) the LFs of the young populations in the cluster region do not show a significant difference to the ones in the reference field.

\section{Reducing background contamination in L16}
\label{sec:tech}

To remove the background contamination in the CMDs of the clusters, L16 used the following technique:

\begin{enumerate}
\item The CMD of the cluster and field region are gridded in several bins/cells.
\item The number of stars within each grid cell in the field region CMD are counted.
\item  In the cluster region CMD, the same number of stars as in the corresponding grid cell of the field region CMD are randomly removed, accounting for the difference in solid angle (area) between the cluster and field regions.
\item The resulting subset of stars in the cluster CMD is considered the decontaminated CMD of the cluster.
\end{enumerate}

This technique will perform well, i.e. reducing significantly the contribution of field stars, only in well populated grid cells, where Poisson uncertainties are much smaller than the number of available stars. However, if the grid cells are populated with just a few stars, the performance of this technique can be very poor. Another caveat of this method is that grid cells that contain more field stars than cluster stars end up with negative values, which are not taken into account during  the analysis. This is the case of the regions of the CMDs that host the young population of the LMC/SMC field (i.e. these regions contain both positive and negative counts after using this technique).

\section{Performance of L16 background decontamination}
\label{sec:per}
\subsection{Subtracting the field from the field}
\label{sec:test}

To illustrate the flaw of the technique used by L16, we carry out a simple experiment. For this we have randomly assigned the stars of the L16 field region to two subsets, irrespective of their spatial location. The experiment consists of taking these two subsets of stars,
and applying the L16 method to reduce the background using one subset as a primary field and the other as reference/background population. With this experiment one would expect to find very few stars at the end, as we are subtracting populations that are statistically identical. If on the contrary, we find that this test yields significant residuals, one can conclude that the technique used is not adequate.

For these experiments we have used the same grid/binning as L16, i.e. grid cells of magnitude $\times$ colour $=0.5\times0.25$ mag$^2$ ranging from $(B-I)=-2.5$ mag to 3.5 and $B=16$ to 27 mag.

\begin{figure}
\centering
\includegraphics[width= 65mm]{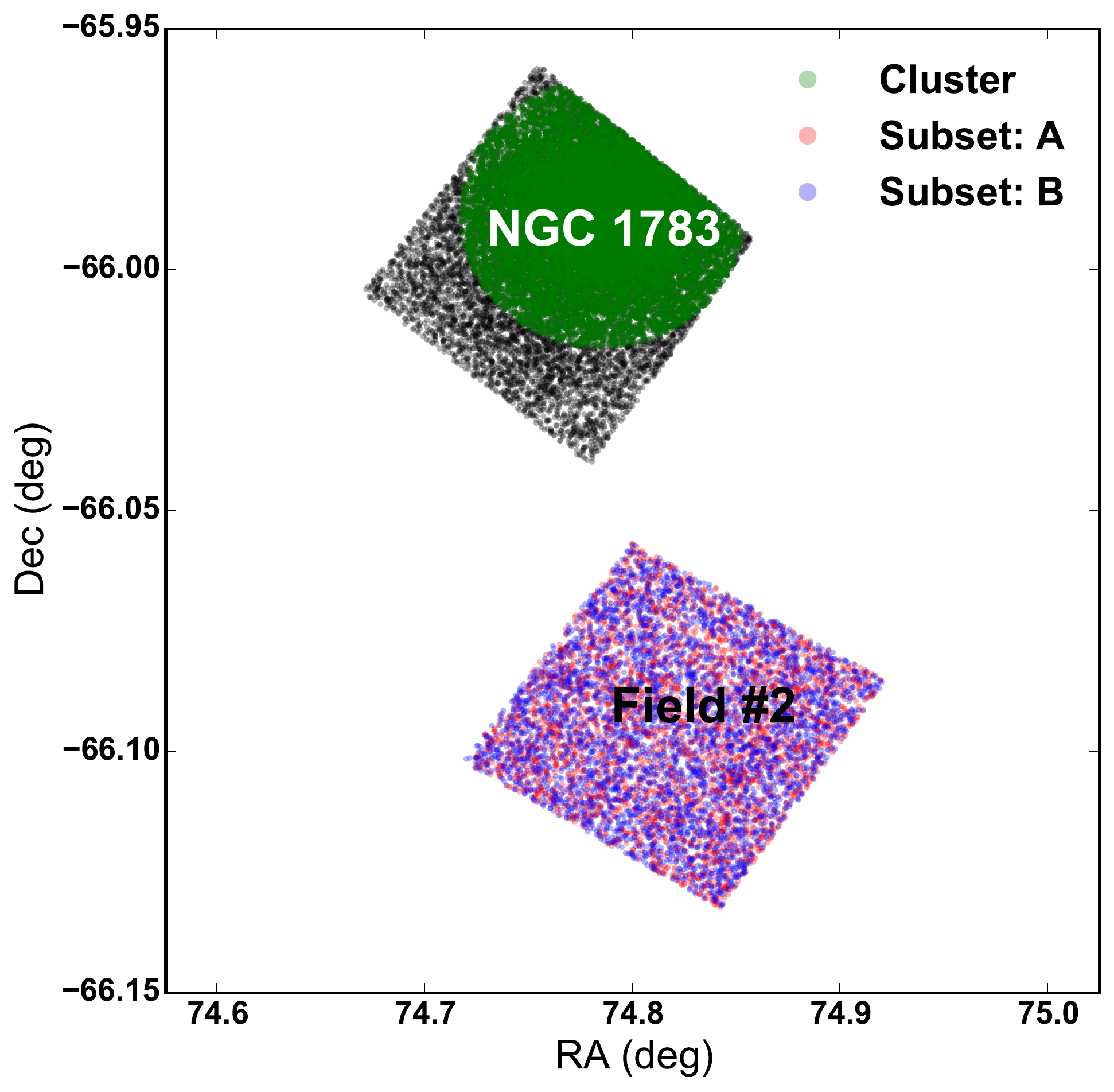}
\caption{Spatial distribution of the cluster and field stars used in our experiments. Red and blue colours show the position of the stars from subset A and B respectively.}
\label{spatial}
\end{figure}

\begin{figure*}
\includegraphics[height= 70mm, width= 34mm]{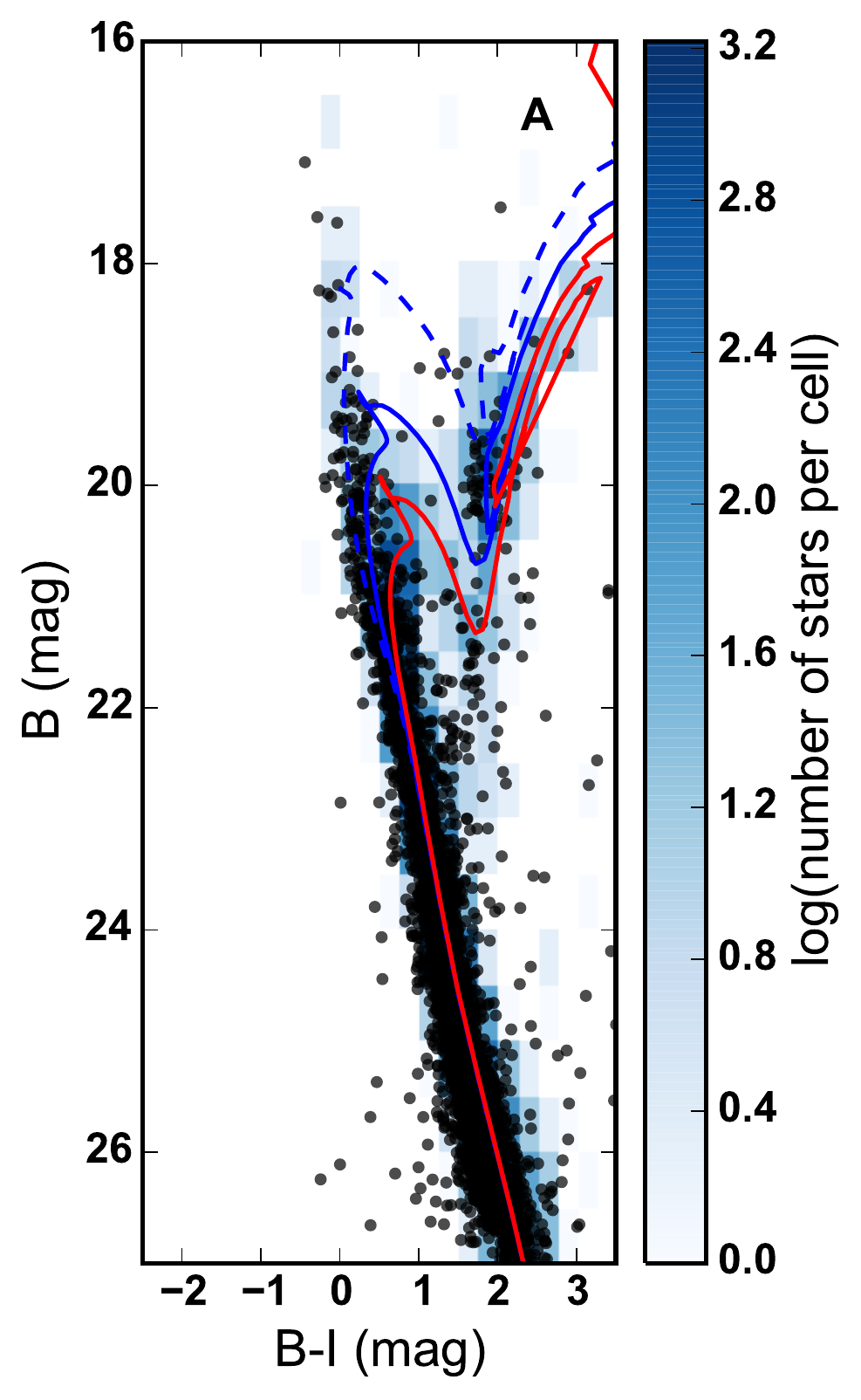}
\includegraphics[height= 70mm, width= 34mm]{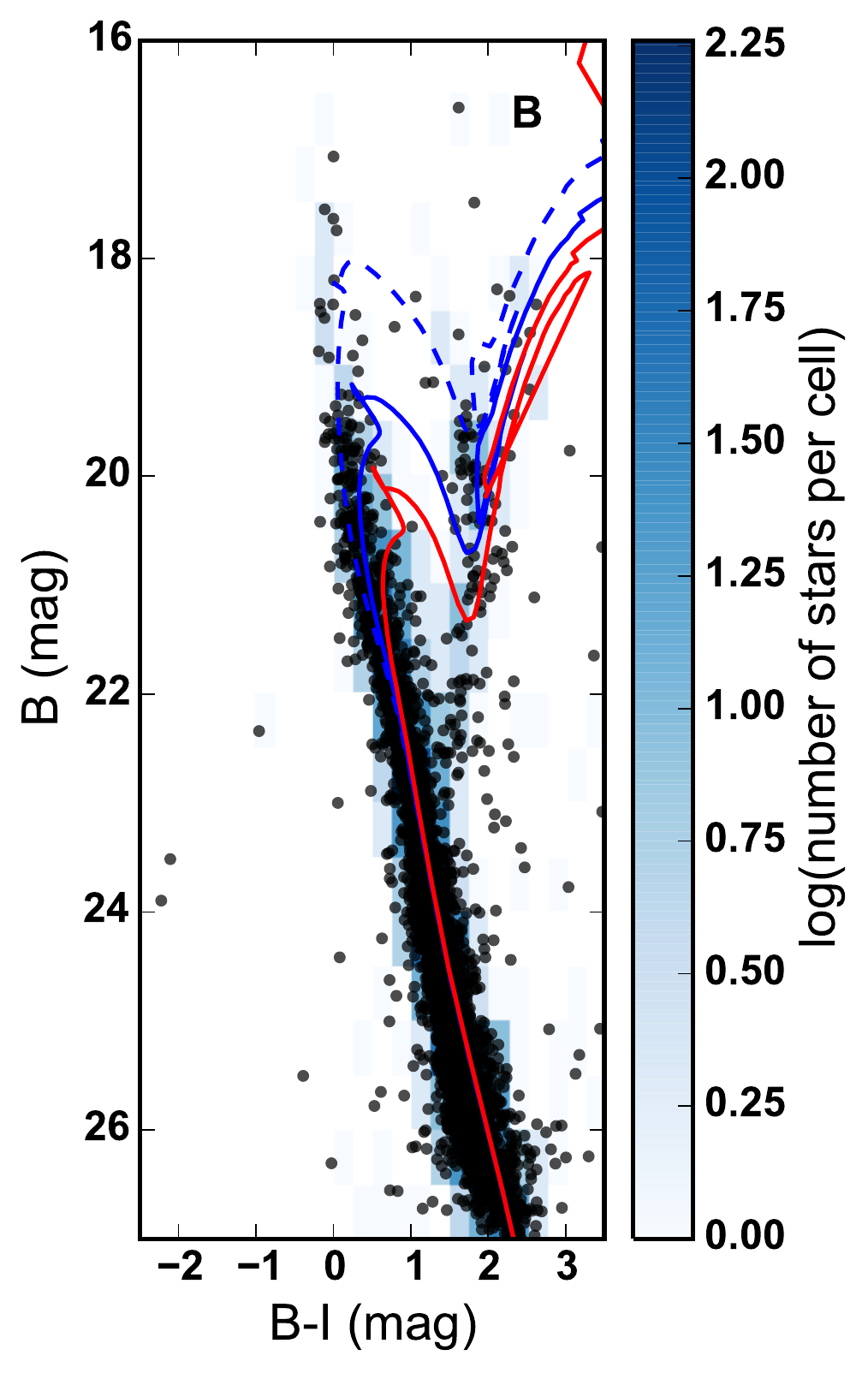}
\includegraphics[height= 70mm, width= 34mm]{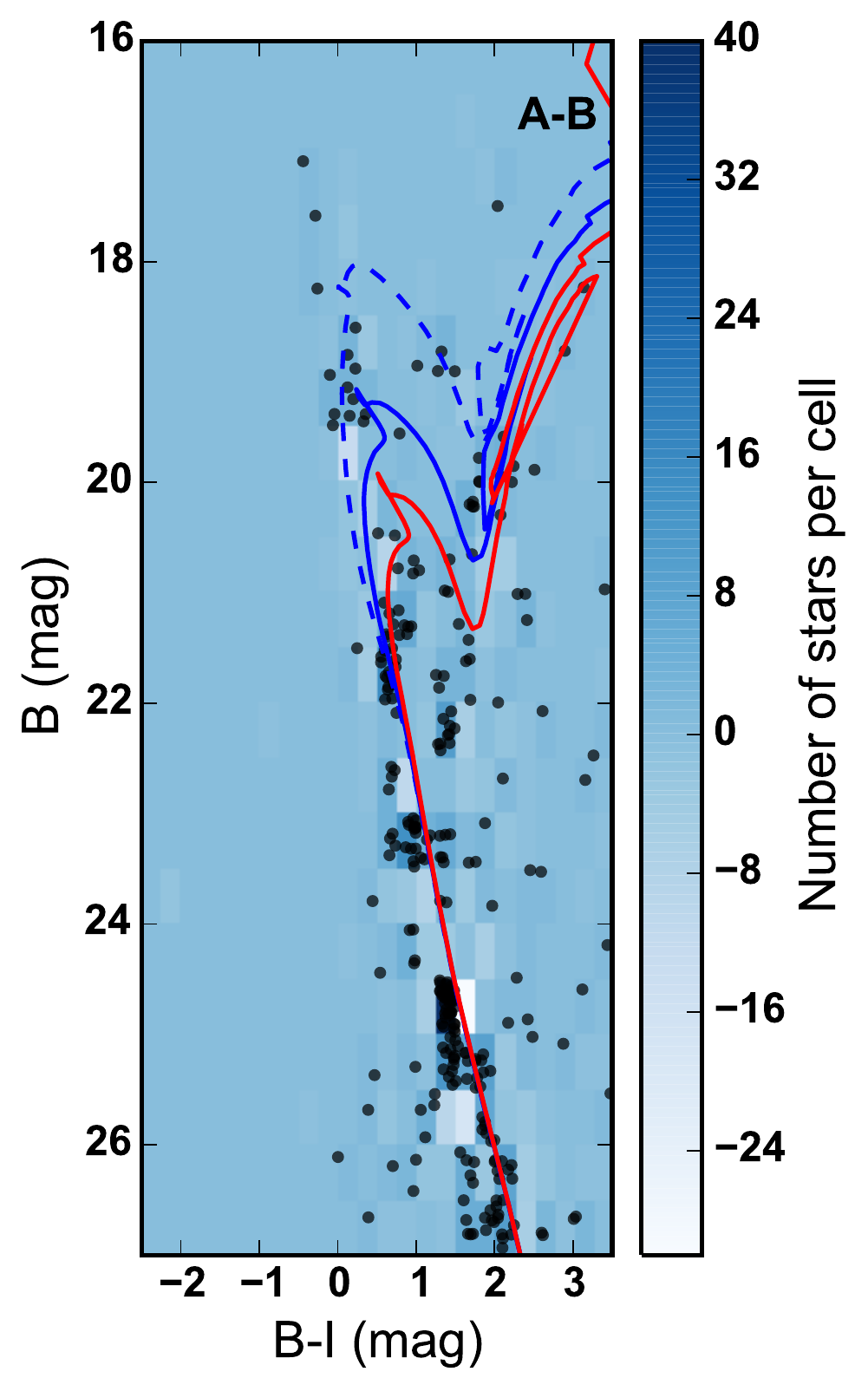}
\includegraphics[height= 70mm, width= 34mm]{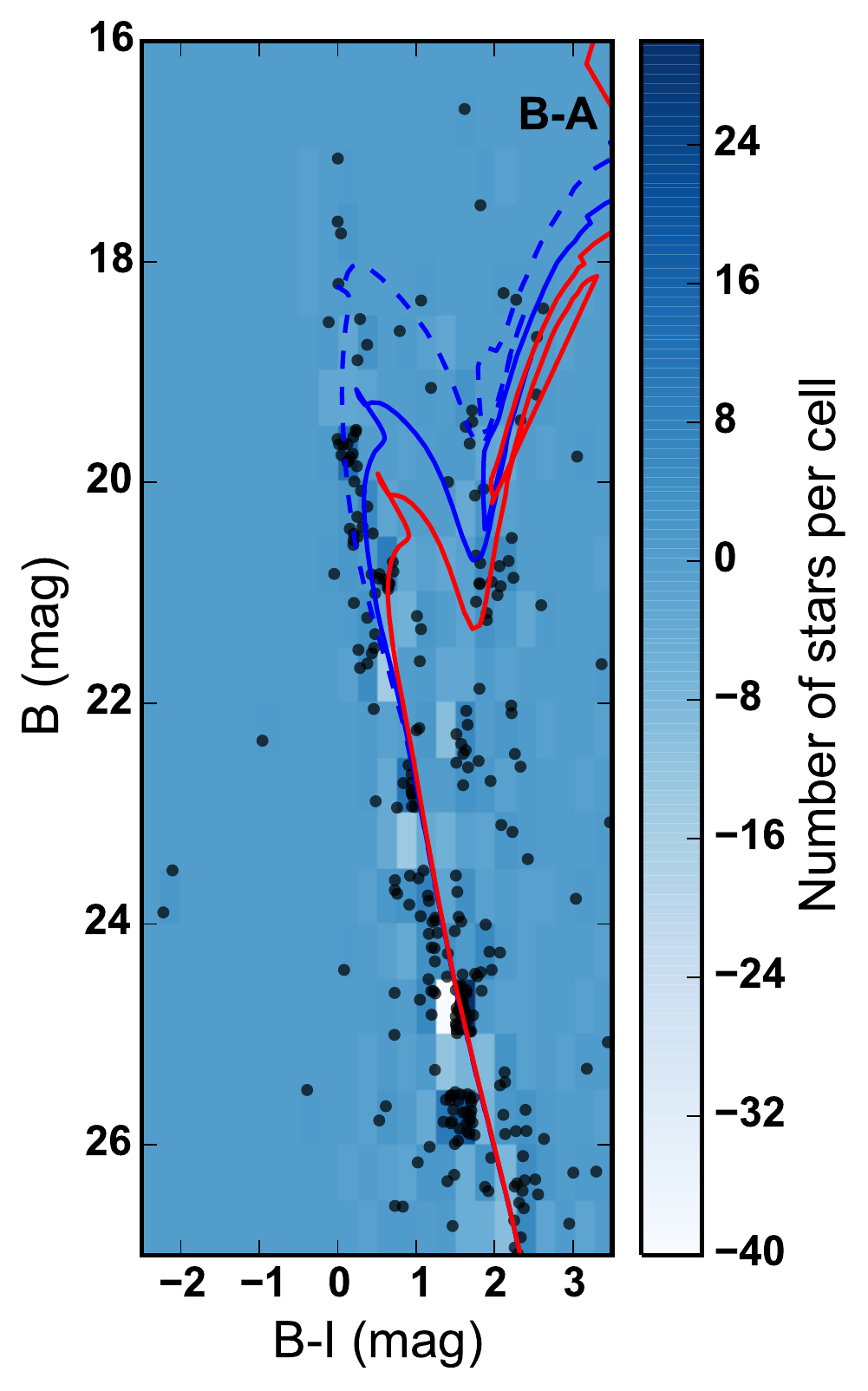}
\includegraphics[height= 70mm, width= 34mm]{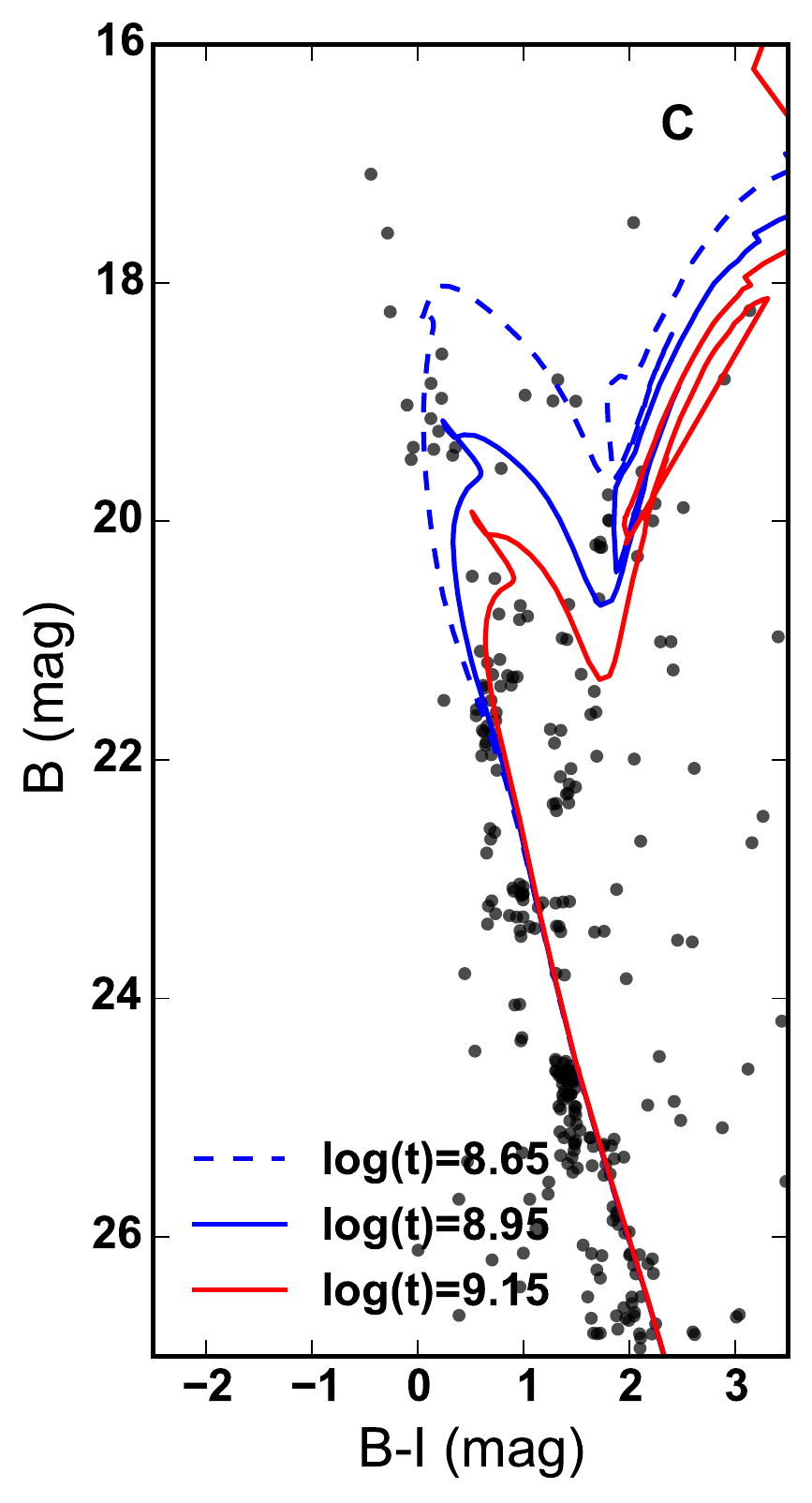}
\caption{CMDs of subset A and B are shown in panels ``A'' and ``B'' respectively. The colour scales in these two panels show the log of the number of stars in each grid element. Panel ``A-B'' shows the residual CMDs obtained when we applied L16's decontamination method assuming that subset A represents the CMD of the stars in the primary field and subset B the reference/background stars. Panel ``B-A'' is similar to panel ``A-B'' but here subset that B was taken as the stars in the primary field and subset A the reference stars instead. The colour scale of panels ``A-B'' and ``B-A'' show the number of stars in each grid cell in linear scale. Panel ``C'' is the same as panel ``A-B'', but without the colour scale. The isochrones in all panels are the same attributed by L16 to the young sequences in their CMD of NGC 1783.}
\label{cmd1783}
\end{figure*}

In Fig. \ref{spatial}, we show the spatial distribution of the subsets of stars taken for our experiment. The subsets A and B, have the same number of stars and were chosen randomly. Both are distributed uniformly across the field region next to the cluster NGC 1783. The CMDs of both subsets are shown respectively in panels ``A'' and ``B'' of Fig. \ref{cmd1783}. The CMDs are virtually identical. As noted in \S \ref{sec:dens}, the young populations found in L16 are already present in the field region. 
We show the \cite{Marigo08} isochrones attributed by L16 to each population for reference. The colour scale represents the log of the number of stars in each grid element.

We then proceed with our experiment, assuming that the subset A contains stars that belong to the primary field, and subset B stars form the reference/background region. After applying the technique used by L16, we were left with the CMD shown in panel ``A-B'' of Fig. \ref{cmd1783}.

In this ``residual'' CMD we note that this technique was efficient in removing most of the stars along the field's MS, i.e. stars with $B>22$ mag. In these regions we are left with $\pm \sim15$ stars/cell -- the negative values correspond to grid cells where there are not enough stars to subtract, i.e. fewer stars in the grid cells of the primary than in the reference/background region, more on this in \S \ref{sec:snr}. When we compare these values to the original number of stars in these grid cells $\sim200$ (i.e. before applying the decontamination technique), we are left with about $\sim8\%$ of the original number of stars in this region of the CMD.


On the other hand, there are significantly fewer stars close to the turn-off of the young populations ($B<20$ mag) with $\sim5$ stars/cell in comparison to $\sim200$ stars/cell in the fainter part of the CMD. The problem with the decontamination method used by L16 is particularly evident here. After ``decontaminating" this part of the CMD, the number of stars per grid cell does not change much -- there are still $\pm5$ stars/cell. In other words, the residuals after decontamination are comparable to the original population due to low number statistics.

The same result is obtained if we take as primary-field stars the background population B and use background population A as reference-field stars, as shown in panel ``B-A'' of the same figure. In this technique, only grid cells with positive counts are analysed, i.e. grid cells resulting in negative counts are ignored, resulting in a bias in the analysis.

We have carried out similar tests, defining different subsets of stars A and B. Also, we have divided ``Field \#2" (cf. Fig. \ref{spatial}) in different spatial subsets, and carried out the same experiments in each of them. All our tests show the same results: the residuals of the young populations are present after the statistical decontamination.

We find the same results for NGC 1696 and NGC 411, the other clusters studied by L16. 
The presence of these residuals after our experiments also calls into question the association of the young populations with these clusters, as they show the ``noise'' leftover after decontaminating a CMD with this technique.


\subsection{Decontaminating the cluster CMD}
\label{sec:test2}

\begin{figure}
\centering
\includegraphics[width= 54mm,height=72mm]{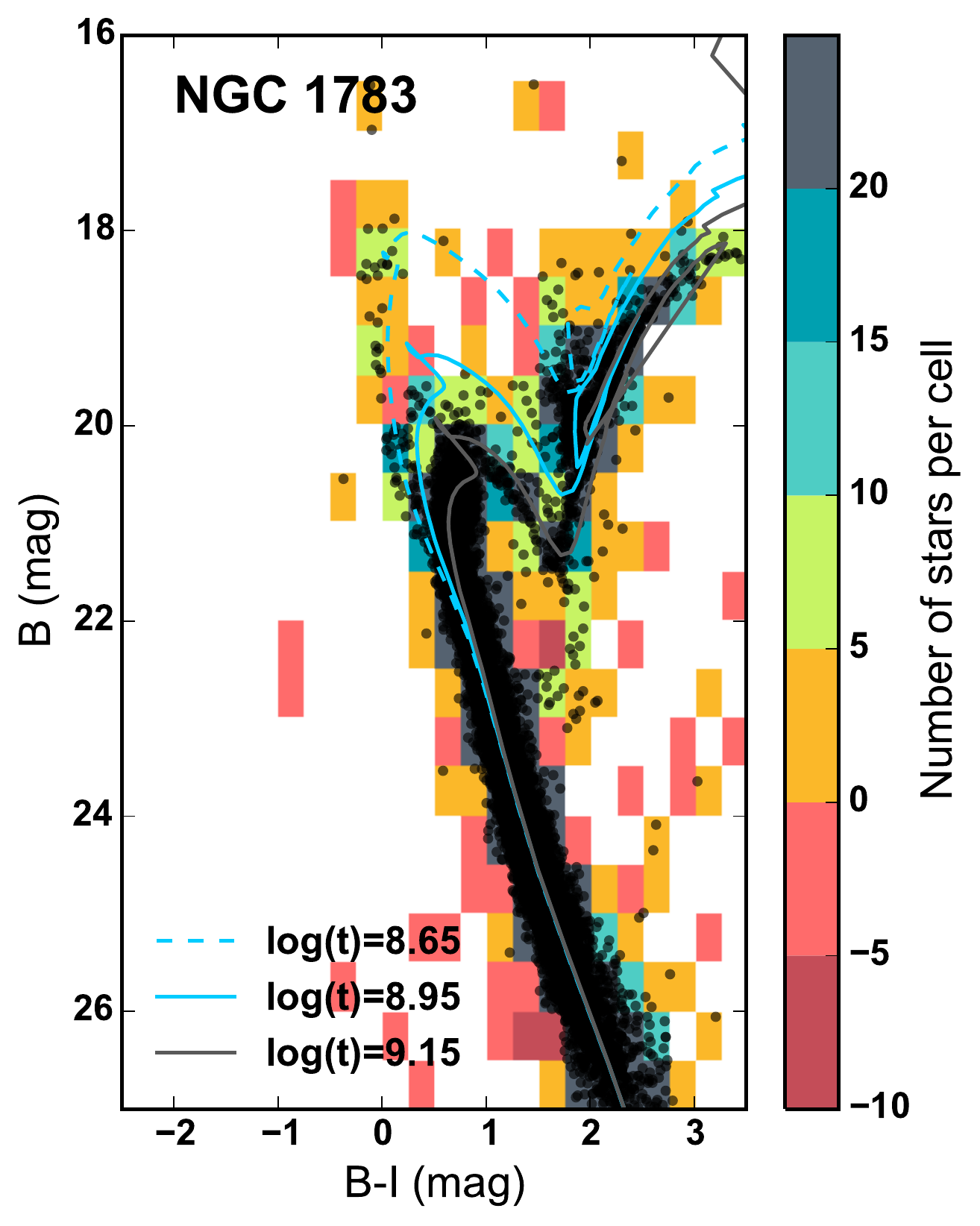}
\caption{CMD of NGC 1783 produced by L16's decontamination technique. In colour, we represent the number of star/cell. The negative values represent the grid cells where there were more stars in the reference field than in the cluster region.}
\label{counts}
\end{figure}

We have applied L16's method to decontaminate the CMD of the clusters using the same reference field as L16 and the same CMD grids. The results obtained for NGC 1783 are shown in Fig. \ref{counts}.

The red cells in this figure represent ``negative cells'', i.e. the cells where the number of stars of the reference field CMD was greater than the number of stars in the cluster region CMD. This feature, ``the negative stars'', is intrinsic to L16's decontamination technique and is a direct consequence of the Poisson regime, where the error in the number of counts is of the order of the number of counts itself. In Fig. \ref{cmd1783}, this is clearly observed when we compare panels ``A-B'' and ``B-A'', as the ``negative stars'' that are missing in one CMD are observed in the other. As a consequence of this effect we have the gaps in the CMDs observed in the panels ``A-B'', ``B-A'' and ``C'' of Fig. \ref{cmd1783}. In Fig. \ref{counts}, this is responsible for the gaps in the faint end of the MS of NGC 1783. 
 Similar effects were readily observed in the first figure of L16, were the younger populations following the isochrones (square symbols) do not show stars below a certain threshold in $B$ magnitude. This gap is not expected if these younger populations follow conventional initial mass functions, so it is likely that the ``gaps'' simply represent ``negative star'' counts.

\section{Significance of the L16 detections}
\label{sec:snr}

We used the method outlined in \cite{Knoetig14} to calculate the probability of the stars in the on-cluster CMD to belong to the background/field and also to quantify the significance of the detections. The solution presented by Knoetig assumes that the on- and off-cluster cell counts follow a Poissonian distribution, and this method has the advantage to be applicable to cells with small and large number of counts, which is ideal in our case, as the number of counts in a cell changes significantly across the CMDs as shown in \S\ref{sec:per}. We refer the reader interested in the details of this method to \cite{Knoetig14}, as the discussion of such a rigorous analysis escapes the scope of this paper.

This method depends only on three parameters: the number of stars in a cell of the on-cluster CMD, the number of stars in a cell of the off-cluster (background/reference) CMD, and the ratio of exposure times between the on- and off-cluster pointings, $\alpha$. We have adopted $\alpha=1$, as is adequate to the regions of the CMD we are interested in, i.e. the ones hosting the young populations, however this assumption need not necessarily be correct for the faint end of the CMD where incompleteness might play a role due to the different exposure times (minimum and maximum exposure times between the on- and off-cluster pointings are 680 s and 720 s respectively).

\begin{figure}
\centering
\includegraphics[width= 54mm,height=72mm]{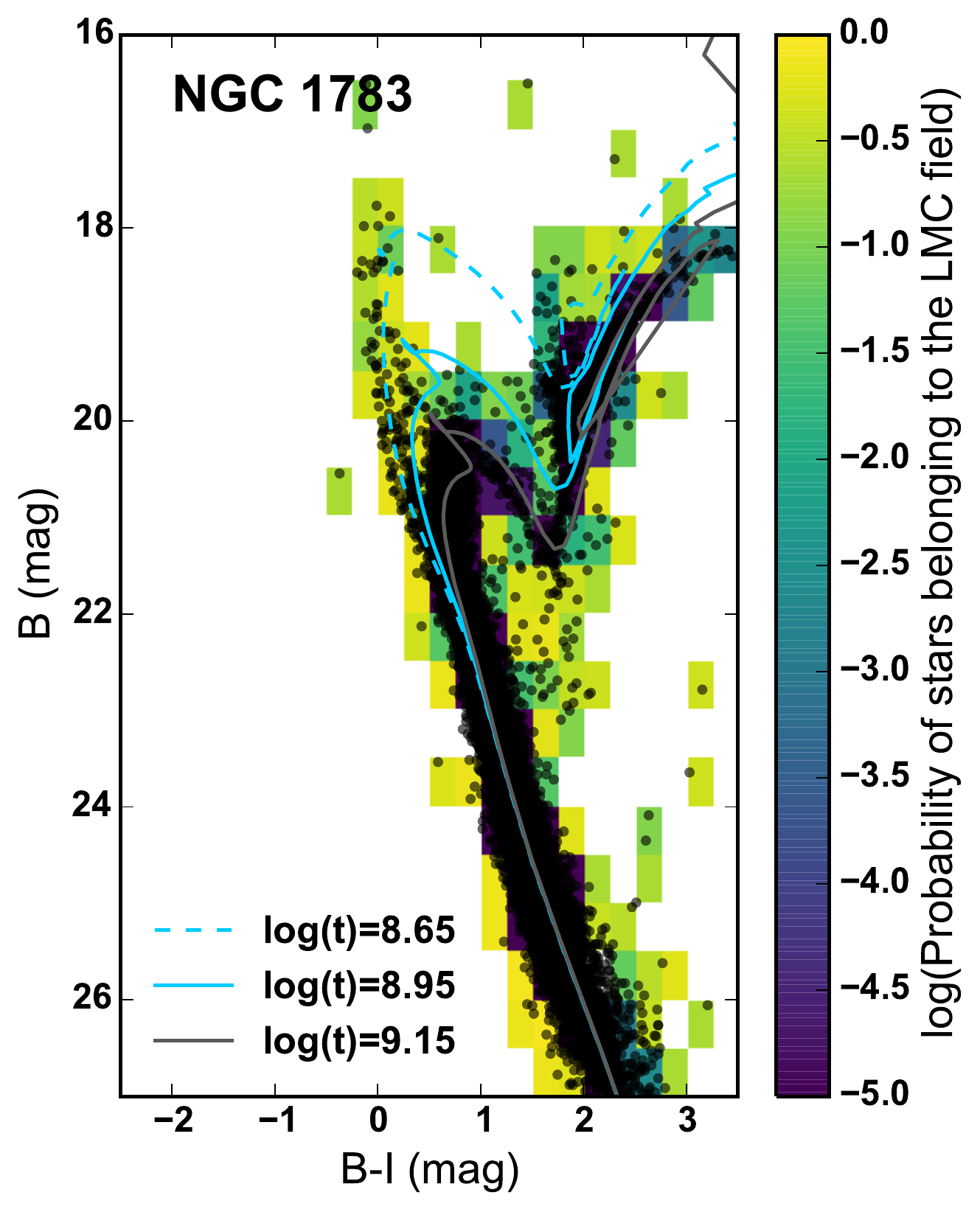}
\caption{On-cluster CMD of NGC 1783, before decontamination. The colour scale represents the log of the probability of the stars in a given cell to belong to the LMC field. The main population has very low probabilities while the young population is very likely to be members of the LMC-field.}
\label{prob}
\end{figure}

We use equation (23) in \cite{Knoetig14}, to calculate the probability of the cell counts in the on-cluster CMD to be only due to background/field stars. In Fig. \ref{prob}, we show the raw CMD of NGC 1783 (i.e. before decontamination), and in colour we represent the (log of the) probability of the stars in the on-cluster region to belong to the background/LMC-field population. Note that all cells that do not belong to the main ($\sim1.5$ Gyr) population, have large ($\sim10-85\%$) probabilities to belong to the LMC field.
On the other hand, the probability that stars along the main cluster sequence belong to the LMC field is vanishingly small, less than $10^{-4} \%$ and as small as $10^{-31} \%$ along the main sequence\footnote{Note that the colour scale in Fig. \ref{prob} is truncated to highlight the contrast between the young and main population.}.

\begin{figure}
\centering
\includegraphics[width= 54mm,height=72mm]{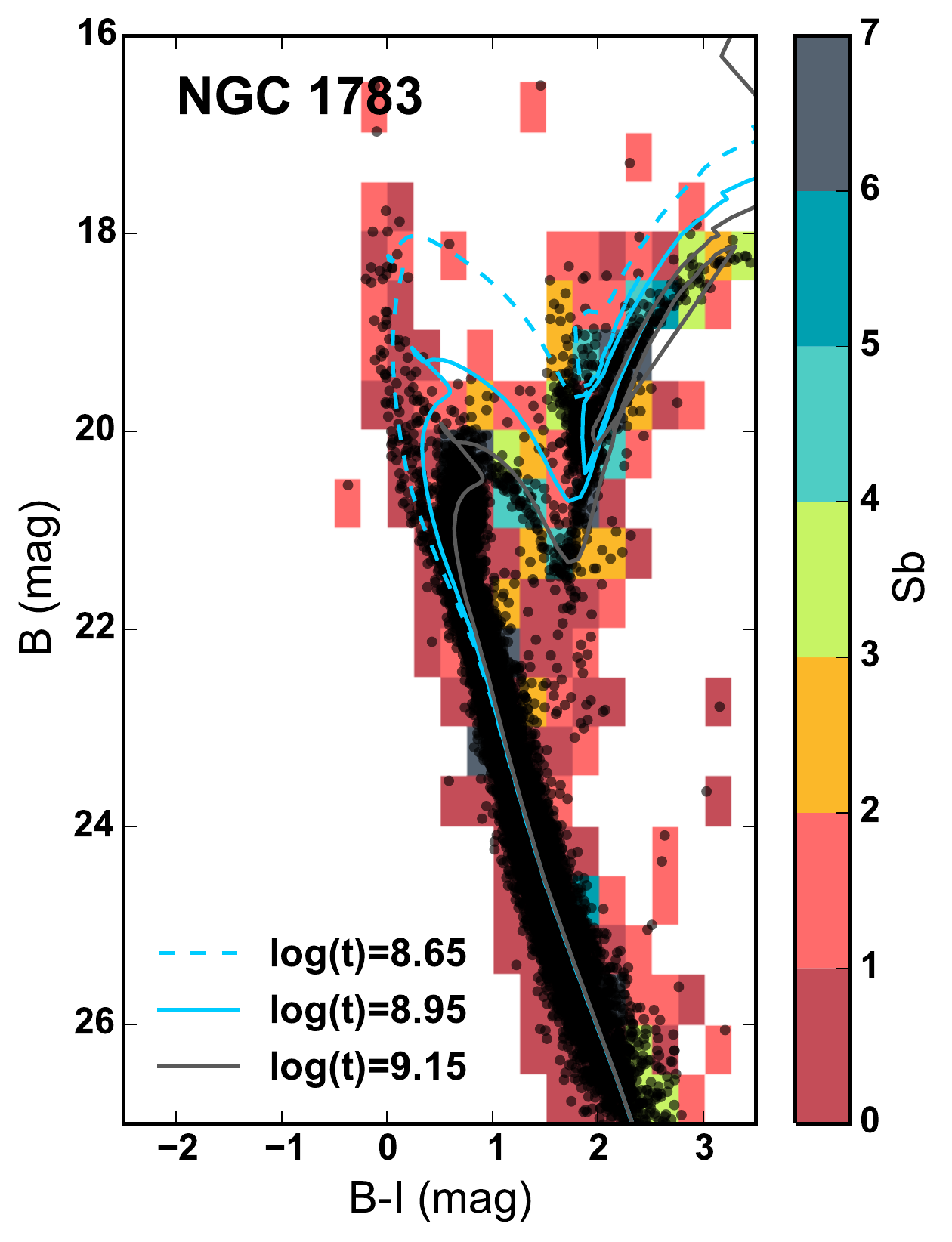}
\caption{On-cluster CMD of NGC 1783, before decontamination. The colour scale represents the significance of the detection in a given cell with respect to the LMC field. The main population is detected at high (>5) significance. On the other hand, the significance of the detection of the young population is minimal, i.e. consistent with the LMC field.}
\label{sig}
\end{figure}

We have also calculated the significance of the detections, $S_b$, defined as ``if the probability were normally distributed, it would correspond to a $S_b$ standard deviation measurement'' (equation (27) in \citealt{Knoetig14}). In other words, it quantifies how much, in standard deviation ($\sigma$) units, the number of stars in a CMD cells differs from the LMC field population. Figure \ref{sig} shows the significance $S_b$ of the detection of stars in a given cell with respect to the LMC field for NGC 1783. The young populations are detected with low (0 -- 2) significance, while the main population is detected at high (>5) significance.



The probability of the young populations to belong to the background/field population and significance of the detections of the young population with respect to the reference field, calculated in this section, also suggest that the young populations from the cluster CMDs belong to the surrounding field.

\section{Spatial distribution of the populations}
\label{sec:spatial}

\begin{figure}
\centering
\includegraphics[width= 72mm]{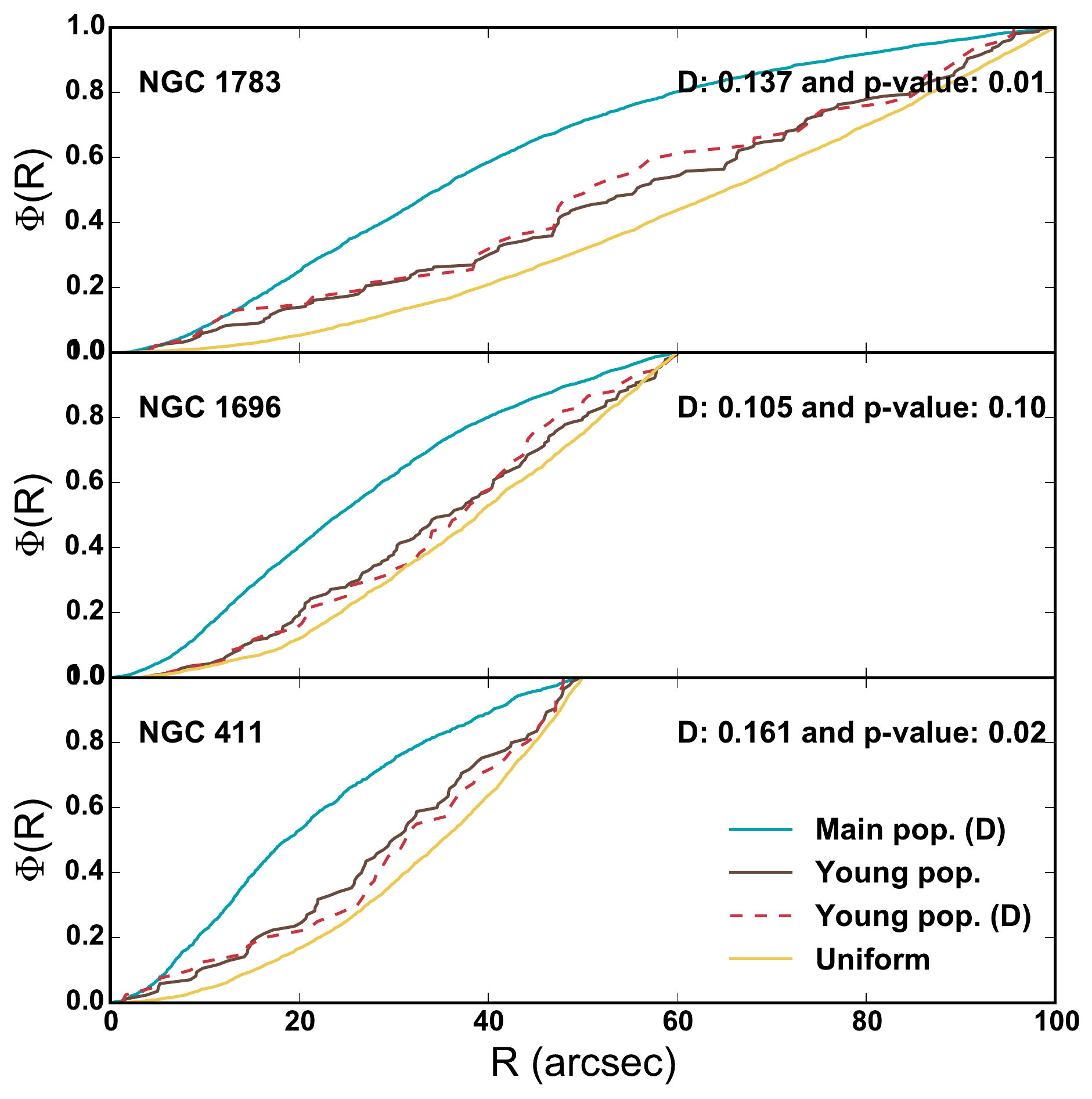}
\caption{Radial profiles of the populations within the clusters. Blue lines: main intermediate-age population after CMD decontamination. Brown and pink lines: young populations, before and after applying L16's decontamination method. Yellow line: synthetic stars distributed uniformly across the clusters. The radial profile of the young populations in these clusters is significantly less centrally concentrated than the main population.}
\label{RPs}
\end{figure}

Finally, here we analyse the spatial distributions of young populations in these clusters. In Fig. \ref{RPs}, we have the radial profiles of the main population, i.e. intermediate-age population, and the young populations found in the clusters'  region before and after CMD decontamination by L16's method. The main populations were taken from $B\mbox{(mag)}<\{21.25,21.50,22.00\}$ and $B-I\mbox{(mag)}\ge\{0.45,0.39,0.14\}$ in the CMDs for clusters NGC 1783, NGC 1696 and NGC 411 respectively (cf. Fig. \ref{young_old}), and the young populations were selected from the same regions as for the LFs in \S \ref{sec:dens}.

From this figure we see that these young populations seem to be less centrally concentrated with respect to the main population of the clusters, in agreement with L16's findings. For comparison, we have distributed randomly, \emph{in a statistically uniform way across each cluster} (i.e. inner two core radii), 10000 artificial stars.
We then perform a KS test comparing the radial profiles of the artificial stars \emph{(uniformly distributed in space)}, and the radial profiles of the young populations before any decontamination. 
The results of these tests are also shown in Fig. \ref{RPs}. 
We conclude that the radial profile of the young population in NGC 1696 is consistent with the radial profile of a uniform distribution of stars, as expected for field (i.e. background/foreground) stars. On the other hand, while the spatial distribution of young stars within NGC 1783 and 411 is not consistent with an uniform spatial distribution, these young stars are significantly less centrally concentrated than the main population of the cluster. We might not expect the young population to be perfectly described by a uniform population, even if they are (as we argue here) likely members of background, if there is a population gradient across the field.

%

The fact that the young populations found in L16's results, are significantly less centrally concentrated than the main population stars, represents another reason to question the association of these young populations to these clusters. Moreover, the young population in NGC 1696 even shares the radial profile expected for uniformly distributed field stars. This is also in agreement with the results presented in \S\ref{sec:dens}, \S\ref{sec:per} and \S\ref{sec:snr}.

We note that the exact shape of the young sequences depends on the reference field adopted and the choice of bin size. On the other hand, these changes have little to no impact to the main (i.e. intermediate age) population of these clusters.

\section{Summary and conclusions}
\label{sec:conc}

We have re-analysed the CMDs of three intermediate-age LMC/SMC star clusters that have been recently claimed to host new generations of stars by L16. Using the same data as L16, we have shown that these young stellar populations belong to the field population of the LMC/SMC. Our experiments have shown that an insufficient background subtraction resulted in these young populations remaining in the clusters' CMDs. We conclude that L16 results are not evidence that these clusters host new generations of stars. This is consistent with previous studies that have looked for, but have not found evidence of multiple epochs of star-formation within young and intermediate-age clusters, like: \cite{Bastian13,Li14,cz14,cz16a,Niederhofer15}.

More sophisticated methods exist to address the issue of field contamination. In the same class as the method adopted by L16 there are applications that properly address the issues of bin edges and placements by taking into account magnitude and colour uncertainties (e.g. \citealt{Kerber02,Balbinot10}). However, more robust methods can be found that adopt an ``unbinned'' approach in a matched-filter framework. Implementations of the latter methods are widely spread across the Local Group dwarf galaxy and stellar stream communities (e.g. \citealt{Martin08,Bechtol15}).

Having said this, given that in all cases we would be dealing with populations in the Poisson regime, one needs to be cautious when interpreting any result obtained for such populations.

In L16, the authors proposed that these clusters were able to accrete and retain gas from their surroundings (adopting the models of \citealt{CS11}), which subsequently spawned a new generation of stars.  Gas accretion and the gas content of star clusters have been studied in several different contexts. So far the evidence points to clusters becoming gas free at very early ages, in most  cases just after a few Myr, e.g. \cite{Hollyhead15}. 
Other studies have shown that clusters remain gas free, even if they are, in principle (based on escape velocity arguments), massive enough to accrete and retain gas from the surrounding, up to very old ages (e.g. \citealt{BS14,BHCZ14,cz15,cz16a,McDonald15}). All this suggests that stellar clusters are extremely inefficient holding onto gas within them. Perhaps this is the reason why, to date, we have not found compelling evidence for multiple stellar generations within clusters.

We note that L16 found that the ``young'' stars in each of the clusters were significantly less centrally concentrated than the main stellar population, in contrast with expectations of models that invoke multiple epochs of star formation in clusters (e.g. \citealt{CS11}). However, if these stars are field contaminants, as argued in the current work, the similar less centrally concentrated distribution is consistent with a field population that was not fully subtracted. 

Finally, L16 adopt He enriched isochrones to explain the younger generation of stars in two of the three clusters, as standard isochrones did not fit the data (for the adopted distance and extinction of the cluster).  Why material accreted from the ISM would be He enriched (and why we do not see stars forming He enriched in the field or clusters/associations today) is left unanswered.

\section{Acknowledgements}

We thank the referee for his/her useful comments. We are grateful to Chengyuan Li for kindly providing his photometry and catalogs. We thank Ricardo Schiavon and Fred Beaujean for helpful discussions. NB is partially funded by a Royal Society University Research Fellowship and an European Research Council (ERC) Consolidator Grant (Multi-Pop - 646928). MG acknowledges financial support from the Royal Society (University Research Fellowship), and MG and EB thank the ERC (ERC-StG - 335936, CLUSTERS) for support.
\bibliographystyle{mnras}
\bibliography{cz16c}

\bsp	
\label{lastpage}
\end{document}